\documentclass[aps,prc,preprint,groupedaddress,12pt]{revtex4-1}

\usepackage{natbib}
\usepackage{amsmath}
\usepackage{graphicx}
\usepackage{hyperref}
\usepackage{multirow}
\usepackage{bm}
\begin{document}

\preprint{\vbox{\hbox{JLAB-THY-16-2269}}}

\title{Factorization Breaking of $A^T_d$ for polarized deuteron targets in a relativistic framework} 

\author{Sabine Jeschonnek$^{(1)}$ and J. W. Van Orden$^{(2,3)}$}

\affiliation{\small \sl (1) The Ohio State University, Physics
Department, Lima, OH 45804\\
(2) Department of Physics, Old Dominion University, Norfolk, VA
23529\\and\\ (3) Jefferson Lab\footnote{Notice: Authored by Jefferson Science Associates, LLC under U.S. DOE Contract No. DE-AC05-06OR23177. The U.S. Government retains a non-exclusive, paid-up, irrevocable, world-wide license to publish or reproduce this manuscript for U.S. Government purposes.}, 12000 Jefferson Avenue, Newport
News, VA 23606
 }

\date{\today}

\begin{abstract}
We discuss the possible factorization of the tensor asymmetry $A^T_d$ measured for polarized deuteron
targets within a relativistic framework. We define a reduced asymmetry and find that factorization
holds only in plane wave impulse approximation and if $p$-waves are neglected. Our numerical results
show a strong factorization breaking once final state interactions are included. We also compare the
$d$-wave content of the wave functions with the size of the factored, reduced asymmetry and find that 
there is no systematic relationship of this quantity to the d-wave probability of the various wave functions.

\end{abstract}

\maketitle

\section{Introduction}

The d-wave contribution to the deuteron has been a matter of interest since it was first determined that such a contribution is responsible in large part for the quadrupole moment of the deuteron. Early calculations of the deuteron wave function including the tensor force needed to produce the $d$-wave admixture resulted in a considerable range of values for the $d$ state probability. Eventually it was argued that the $d$-wave contribution was not a physical observable \cite{Friar1977,Friar1979,Amado:1978zz,Amado:1974xh}.

The argument for this depends on the realization that there are two different distances which need to be used in the analysis of deuteron observables \cite{Friar1979,Amado:1978zz,Amado:1974xh}. 
First it should be noted that the physical observables for a reaction are matrix elements that involve the measurement of the values of the initial and final states at distances that are very large. Here,
very large means very large compared to the size of the region in which the constituents of the deuteron interact to provide the initial and final states where all particles are onshell. 
All quantities that only involve measurement at these large distances are ``outside'' quantities while the details of the quantities within the interaction region are ``inside'' quantities. The ``inside'' quantities generally involve contributions that are either off mass shell or off energy shell depending on the method used to calculate the matrix element. The ``inside'' quantities include wave functions and transition operators which produce the asymptotic final state from the asymptotic initial state. The ``inside'' quantities are not, in general observables. 

Consider the case of the matrix element for deuteron electrodisintegration
\begin{equation}
\left<\bm{p}_1s_1;\bm{p}_2s_2\right|J^\mu(q)\left|\bm{P}\lambda_d\right>\,, \label{eq:current}
\end{equation}
where the final state consists of a proton of momentum $\bm{p}_1$ and spin $s_1$ and a neutron of momentum $\bm{p}_2$ and spin $s_2$. The initial state consists of a deuteron with momentum $\bm{P}$ and helicity $\lambda_d$. The four-vector current operator for four-momentum transfer $q$ is $J^\mu(q)$. In general this contains both one- and two-body contributions. Defining a unitary transformation $U$ which is effective only in the ``inside'' region, the matrix element (\ref{eq:current}) can be rewritten as
\begin{equation}
\left<\bm{p}_1s_1;\bm{p}_2s_2\right|U^{-1}UJ^\mu(q)U^{-1}U\left|\bm{P}\lambda_d\right>=\overline{\left<\bm{p}_1s_1;\bm{p}_2s_2\right|J^\mu(q)\left|\bm{P}\lambda_d\right>}\,.
\end{equation}
The value of the matrix element is not changed by the unitary transformation, but the initial and final states and the current operator are each modified.

Two examples of the kind of unitary transformation that can be used in this context: the first of these is the use of field redefinition in effective Lagrangians that are used to model the $NN$ interaction and electromagnetic currents \cite{Friar1979,Adam:1997pw}. For example such definitions can change the relative size of pseudoscalar and derivative couplings for the interaction and therefore change the size of the deuteron $d$-state and the current operator.

The second type of unitary transformation is to use the renormalization group \cite{furnstahl_evolve,furnstahl_schwenk_nonobs,furnstahldeu,furnstahl_hammer} to reduce the relative momentum that can contribute to interaction. This acts as a cutoff that limits the range of the tensor force in the calculation of wave functions while transferring higher momentum contributions to the current operator. In fact, almost any unitary transformation that acts only on the ``inside'' region can be used to redistribute strength between the wave functions and current operator. This means that the definitions of wave functions and current operators are not unique and therefore the $d$-state component of the deuteron is not a unique measurable quantity.

The one ``outside'' quantity which can be identified with the $d$-state is the asymptotic  $d$ to $s$ ratio \cite{Amado:1978zz} which measures the relative size of the two contributions outside of the interaction region and can be determined by extrapolating the scattering matrix for nucleon scattering from a tensor-polarized deuteron to an unphysical region.

One possible approach that might allow for the determination of some information about the $d$-state would be if some region of kinematics could be found where the sensitivity to the components of the current operator and to the final state interactions are negligible. This would lead to the factorization of the polarized deuteron cross sections into an effective single nucleon part and various polarized momentum distributions.  Appropriate ratios of the cross section can then be defined where the single-nucleon cross sections cancel and only a ratio of momentum distributions remains.  The objective of this paper is to address the feasibility of such a process using a variety of modern high quality wave functions, a selection of single-nucleon electromagnetic form factors, and two different parameterizations of the final state interactions.
The calculations presented here may provide some guidance for future experiments \cite{Longproposal}.

The paper is organized as follows: first, we briefly review the general formalism necessary to calculate response functions for polarized targets, and the definitions of the asymmetries that can be measured for polarized
targets and beams. Then, we carefully discuss under which conditions the factorization of the tensor asymmetry 
may arise in a fully relativistic framework, and which form the asymmetry $A^T_d$ takes when it holds. 
In the next section, we present our numerical results, in a kinematic region relevant to
experiments at Jefferson Lab. We show the influence of the different model inputs on the calculations.
With final state interactions included, factorization breaks. We conclude with a brief summary.

\section{Formalism}

We present a brief review of the formalism for calculating the differential cross section, response functions, and
asymmetries for target polarization. For a complete discussion, the reader is referred to \cite{targetpol}.

\subsection{Differential Cross Section}

\begin{figure}
\centerline{\includegraphics[height=3in]{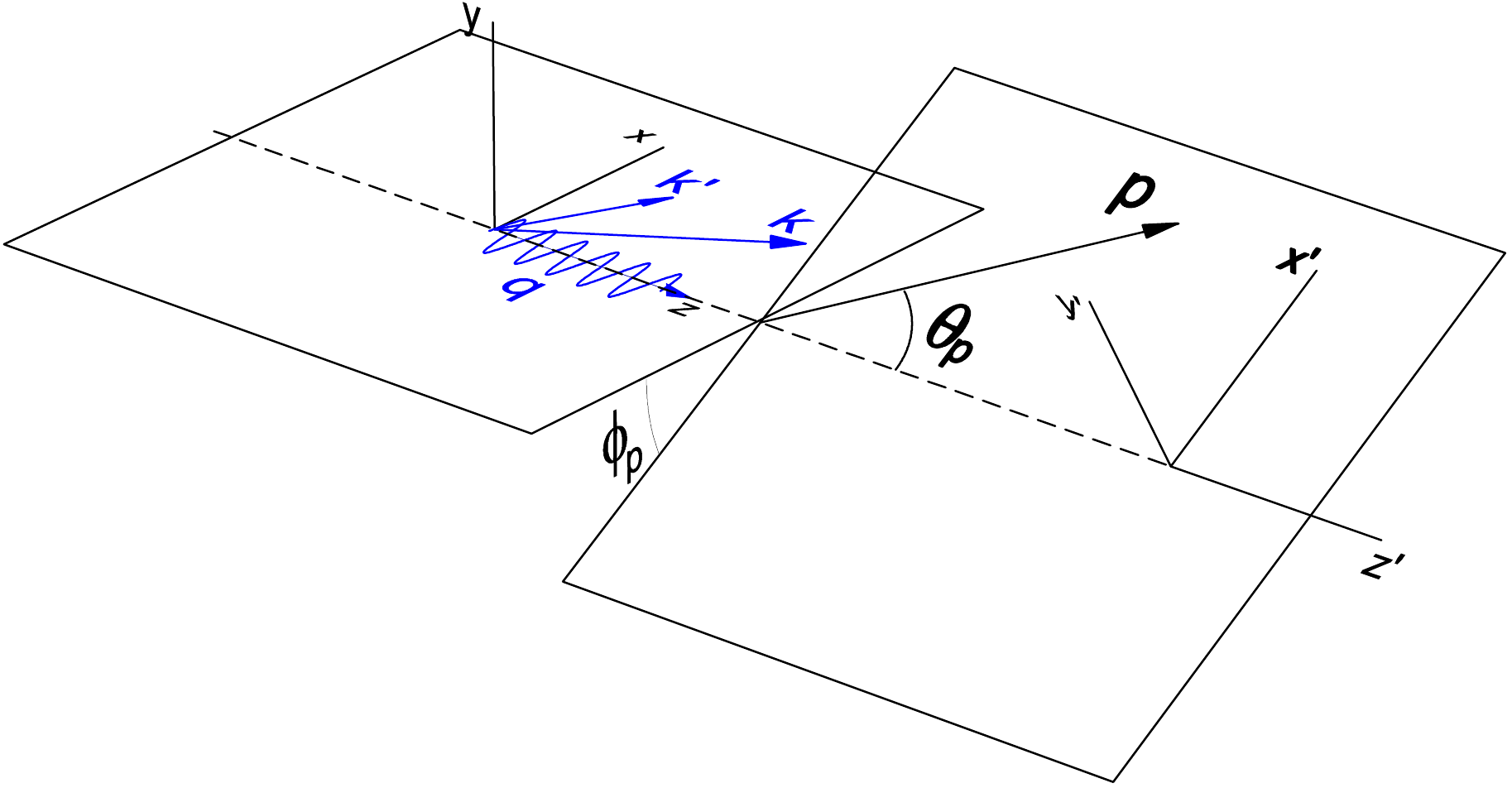}}
\caption{(Color online) Coordinate systems for the $^2H(e,e'p)$ reaction.  $k$ and $k'$ are the initial and final electron four-momenta,
$q$ is the four-momentum of the virtual photon and $p$ is the four-momentum of the final-state proton.  }\label{coordinates}
\end{figure}

The standard coordinate systems used to describe the $D(e,e'p)$ reaction are shown in Fig.\ref{coordinates}.
The initial and final electron momenta $\bm{k}$ and $\bm{k'}$ define the electron scattering plane
and the $xyz$-coordinate system is defined such that the $z$ axis, the quantization axis, lies along
the momentum of the virtual photon $q$ with the $x$-axis in the electron scattering plane and the $y$-axis
perpendicular to the plane. The momentum $p$ of the outgoing proton is in general not in this plane and is
located relative to the $xyz$ system by the polar angle $\theta_p$ and the azimuthal angle $\phi_p$.

The general form of the $^2H(e,e'p)$ cross section can be written in the lab frame as
\cite{raskintwd,dmtrgross}

\begin{eqnarray}
\left ( \frac{ d \sigma^5}{d \epsilon' d \Omega_e d \Omega_p} \right
)_h  & = & \frac{m_p \, m_n \, p_p}{8 \pi^3 \, M_d} \,
\sigma_{Mott} \,
f_{rec}^{-1} \,
 \Big[  v_L R_L +   v_T R_T
 + v_{TT} R_{TT} + v_{LT}R_{LT}
  \nonumber \\
& & +  h \,  v_{LT'} R_{LT'}+h\,v_{T'}R_{T'}
\Big] \, , \label{xsdef}
\end{eqnarray}
where $M_d$, $m_p$ and $m_n$  are the masses of the deuteron, proton and neutron,
 $p_p=p_1$ and $\Omega_p$
are the momentum and solid angle of the ejected proton, $\epsilon'$ is the
energy of the detected electron and $\Omega_e$ is its solid angle, with
$h=\pm 1$  for positive and negative electron helicity. The Mott cross
section is
\begin{equation}
\sigma_{Mott} = \left ( \frac{ \alpha \cos(\theta_e/2)} {2
\varepsilon \sin ^2(\theta_e/2)} \right )^2
\end{equation}
and the recoil factor is given by
\begin{equation}
f_{rec} = \left| 1+ \frac{\omega p_p - E_p q \cos \theta_p} {M_d \, p_p}
\right| \, . \label{defrecoil}
\end{equation}
The hadronic tensor for scattering from polarized deuterons is defined as

\begin{equation}
W^{\mu\nu}(D)=\sum_{s_1,s_2,\lambda_d,\lambda'_d}
\left<\bm{p}_1s_1;\bm{p}_2s_2;(-)\right|J^\mu\left|\bm{P}\lambda'_d\right>^*
\left<\bm{p}_1s_1;\bm{p}_2s_2;(-)\right|J^\nu\left|\bm{P}\lambda_d\right>
\rho_{\lambda_d\lambda'_d}\label{eq:hadron_tensor}
\end{equation}
 The notation $(-)$ in the final state indicates that the state satisfies the boundary conditions appropriate for an ``out'' state.
The deuteron density matrix in the $xyz$-frame is
\begin{equation}
\bm{\rho}=\frac{1}{3}\left(
\begin{array}{ccc}
1+\sqrt{\frac{3}{2}}\,T_{10}+\frac{1}{\sqrt{2}}T_{20}
& -\sqrt{\frac{3}{2}}(T^*_{11}+T^*_{21})
& \sqrt{3}\,T^*_{22}\\
-\sqrt{\frac{3}{2}}(T_{11}+T_{21})
& 1-\sqrt{2}\,T_{20}
& -\sqrt{\frac{3}{2}}(T^*_{11}-T^*_{21})\\
\sqrt{3}\,T^*_{22}
& -\sqrt{\frac{3}{2}}(T_{11}-T_{21})
& 1-\sqrt{\frac{3}{2}}\,T_{10}+\frac{1}{\sqrt{2}}T_{20}
\end{array}\right)
\end{equation}
and the set of tensor polarization coefficients is defined as
\begin{equation}
D=\left\{ U,T_{10},T_{11},T_{20},T_{21},T_{22}\right\}
\end{equation}
with $U$ designating the contribution from unpolarized deuterons. The kinematic factors $v_i$ and the polarized response functions are listed in Appendix \ref{app:kinematic} for convenience.

The expressions above assume that the deuteron target is polarized along the direction of the three momentum transfer $\bm{q}$ which is defined as the z-axis as defined in Fig. \ref{coordinates}. However, since this would require realignment of the target polarization for each value of $\bm{q}$, experiments are performed with the deuteron polarization generally fixed along the direction of the electron beam defined by $\bm{k}$. This involves a right-handed rotation of the deuteron density matrix about the y-axis through the angle between $\bm{q}$ and $\bm{k}$ denoted by $\theta_{kq}$ resulting in a new set of polarization coefficients $\widetilde{T}_{JM}$. This rotation is described in detail in Appendix \ref{app:rotations}. Equation (\ref{eq:T_Ttilde}) relates $T_{JM}$ to $\widetilde{T}_{JM}$.

\subsection{Asymmetries}

Conventionally, target polarization in deuteron electrodisintegration is measured in terms of four single asymmetries
\begin{eqnarray}
\label{asymdef}
A^V_d&=&\frac{v_L R_L(\widetilde{T}_{10})+v_T R_T(\widetilde{T}_{10})+v_{TT}R_{TT}(\widetilde{T}_{10})+v_{LT} R_{LT}(\widetilde{T}_{10})}{\widetilde{T}_{10}\Sigma}\nonumber\\
A^T_d&=&\frac{v_L R_L(\widetilde{T}_{20})+v_T R_T(\widetilde{T}_{20})+v_{TT}R_{TT}(\widetilde{T}_{20})+v_{LT} R_{LT}(\widetilde{T}_{20})}{\widetilde{T}_{20}\Sigma}\nonumber\\
A^V_{ed}&=&\frac{v_{LT'}R_{LT'}(\widetilde{T}_{10})
	+v_{T'}R_{T'}(\widetilde{T}_{10})}{\widetilde{T}_{10}\Sigma}\nonumber\\
A^T_{ed}&=&\frac{v_{LT'}R_{LT'}(\widetilde{T}_{20})+v_{T'}R_{T'}(\widetilde{T}_{20})
}{\widetilde{T}_{20}\Sigma}\,
\end{eqnarray}
where
\begin{equation}
\Sigma=v_L R_L(U)+v_T R_T(U)+v_{TT}R_{TT}(U)+v_{LT} R_{LT}(U)\,.
\end{equation}
Here $R_i(\widetilde{T}_{10})$ and $R_i(\widetilde{T}_{20})$ denote the response functions where only $\widetilde{T}_{10}$ is nonzero or only $\widetilde{T}_{20}$ is nonzero. $R_i(U)$ denotes the unpolarized response functions. In (\ref{asymdef}) the superscripts $V$ and $T$ refer to vector and tensor polarizations. The subscript $d$ indicates that all of these asymmetries are defined for polarized deuterons.  The subscript $e$ denotes the case where the electron beam is also polarized.

\subsection{Factorization}
\label{subsecfac}

\begin{figure}
	\centerline{\includegraphics[width=5in]{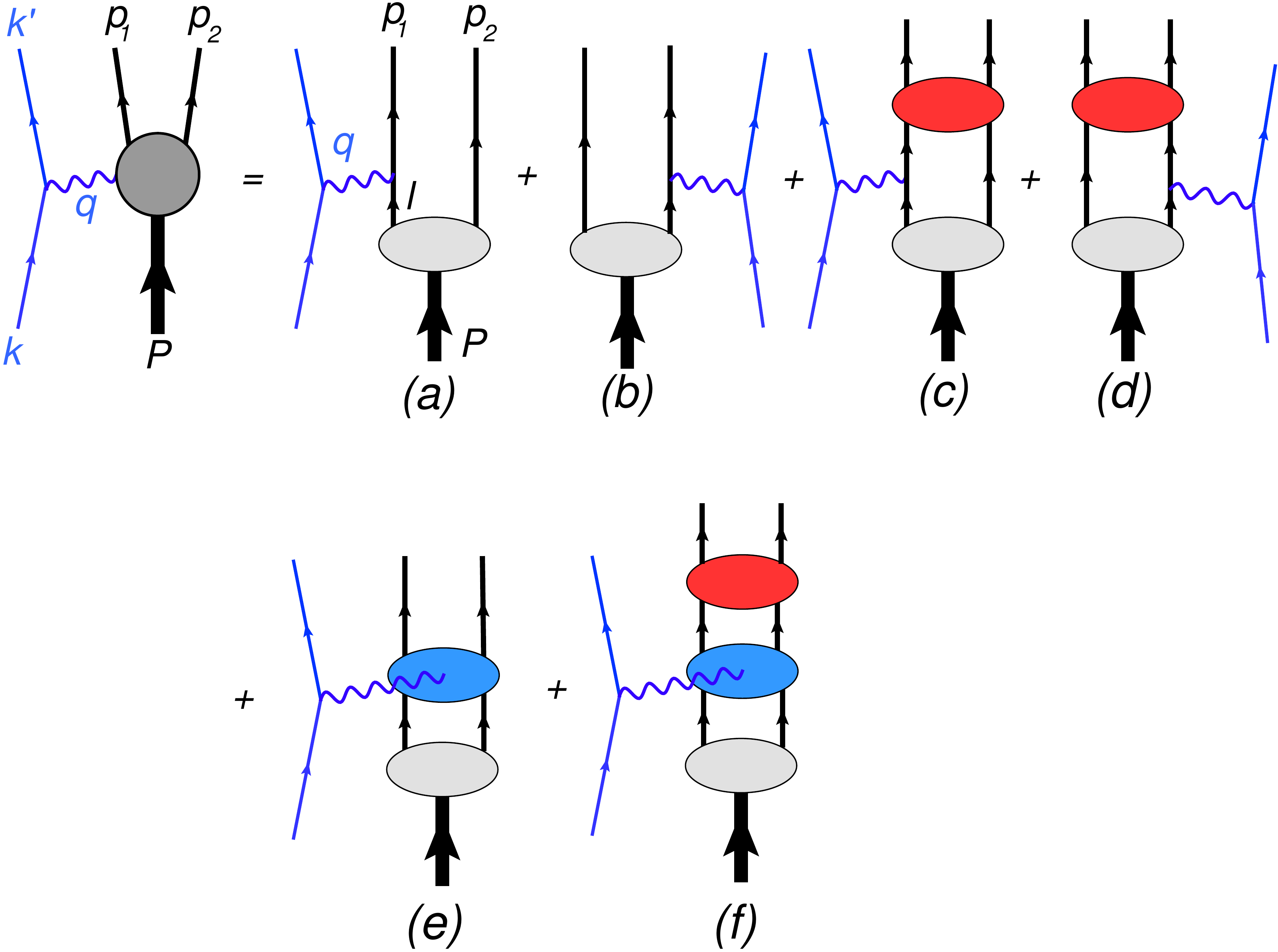}}
	\caption{(color online) Feynman diagrams for the impulse
		approximation.  }\label{fig:current}
\end{figure}

The Feynman diagrams representing current matrix element deuteron electrodisintegration for the Bethe-Salpeter equation are shown in Fig. \ref{fig:current}. Figures \ref{fig:current}(a) and (b) have plane wave (PW) final states while Figs.\ref{fig:current}(c) and (d) include final state interactions (FSI). Figures \ref{fig:current}(e) and (f) contain two-body currents with plane wave final state and FSI respectively.

Any attempt to evaluate the effects of $d$-state contributions to the deuteron wave function require that the cross section can be factored into an effective single-nucleon cross section and a momentum distribution. This can only occur is Figs. (\ref{fig:current})(b)-(f) make negligible contributions to the current matrix element.

\begin{figure}
\centerline{\includegraphics[height=1.5in]{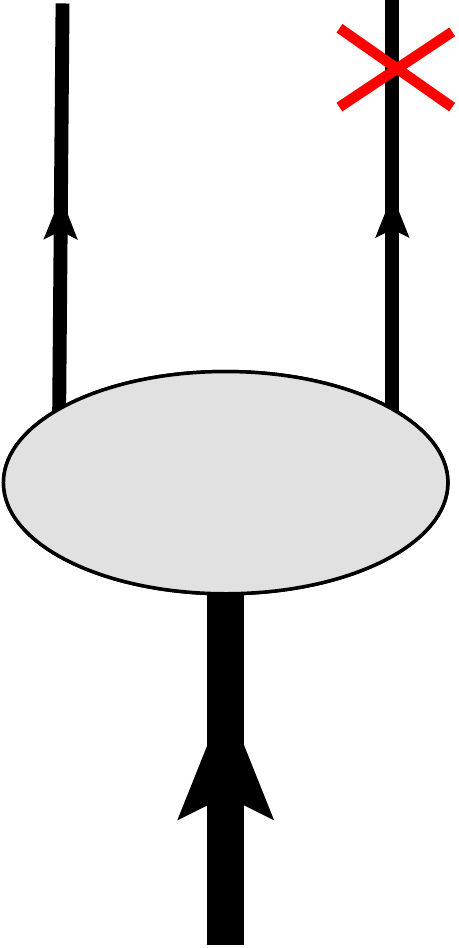}}
\caption{(color online) Feynman diagram representing the half-offshell vertex function.}\label{fig:vertex}
\end{figure}

The plane wave contribution to the current matrix element represented by Fig. \ref{fig:current}(a) can then be written as
\begin{equation}
\left<\bm{p}_1s_1;\bm{p}_2s_2\right|J^\mu_{(1)}\left|\bm{P}\lambda_d\right>_a=
-\bar{u}(\bm{p}_1,s_1)\Gamma^\mu(q)G_0(P-p_2)
\Gamma^T_{\lambda_d}(p_2,P)\bar{u}^T(\bm{p}_2,s_2)\,,
\end{equation}
where the nucleon propagator is
\begin{equation}
G_0(p)=\frac{\gamma\cdot p+m_N}{m_N^2-p^2}
\end{equation}
and the one-body nucleon electromagnetic current operator is chosen to be of the Dirac-plus-Pauli form
\begin{equation}
\Gamma^\mu(q)=F_1(Q^2)\gamma^\mu+\frac{F_2(Q^2)}{2m_N}i\sigma^{\mu\nu}q_\nu\,.
\end{equation}
The deuteron vertex function with particle 2 onshell, as required by Fig. \ref{fig:current}(a) is shown in Fig. \ref{fig:vertex} and is given in general by
\begin{align}
\Gamma_{\lambda_d}(p_2,P)=&\left[ g_1(p_2^2,p_2\cdot
P)\gamma\cdot\xi_{\lambda_d}(P) -g_2(p_2^2,p_2\cdot P)
\frac{p_2\cdot\xi_{\lambda_d}(P)}{m_N}\right.\nonumber\\
&\left.-\left(g_3(p_2^2,p_2\cdot P)\gamma\cdot\xi_{\lambda_d}(P)
-g_4(p_2^2,p_2\cdot
P)\frac{p_2\cdot\xi_{\lambda_d}(P)}{m_N}\right)\frac{\gamma\cdot
(P-p_2)+m}{m_N}\right]C\,.\label{eq:vertex_function}
\end{align}
where $\xi_{\lambda_d}(P)$ is the deuteron polarization four-vector, $C$ is the charge conjugation matrix and the invariant functions $g_i(p_2^2,p_2\cdot P)$ are given by

\begin{align}
g_1(p_2^2,p_2\cdot P)=&\frac{(2 E_{p_R}-M_d) (p_R
   \Psi_3(p_R)-m_N \Psi_4(p_R))}{4
   \sqrt{\pi } p_R}  \\
g_2(p_2^2,p_2\cdot P)=&\frac{m_N (2 E_{p_R}-M_d)
   \left(\sqrt{2} E_{p_R}
   \Psi_1(p_R)-m_N
   \Psi_3(p_R)-p_R
   \Psi_4(p_R)\right)}{4 \sqrt{\pi }
   p_R^2}  \\
g_3(p_2^2,p_2\cdot P)=&-\frac{E_{p_R} m_N \Psi_4(p_R)}{4
   \sqrt{\pi } p_R}  \\
g_4(p_2^2,p_2\cdot P)=&\frac{m_N^2}{4 \sqrt{\pi}
   M_d p_R^2} \Bigl(-2 E_{p_R}^2
   \Psi_3(p_R)+E_{p_R} \left(M_d
   \Psi_3(p_R)+2 \sqrt{2} m_N
   \Psi_1(p_R)\right)\nonumber\\
   &+\sqrt{2} M_d
   (p_R \Psi_2(p_R)-m_N
   \Psi_1(p_R))\Bigr)
\end{align}
where
\begin{equation}
p_R=\sqrt{\frac{(P\cdot p_2)^2}{P^2}-p_2^2}
\end{equation}
is the relative momentum of the nucleons in the rest frame of the deuteron and
\begin{align}
\Psi_1(p_R)&=u(p_R)+\sqrt{2}w(p_R)\\
\Psi_2(p_R)&=-\sqrt{3}v_s(p_R)\\
\Psi_3(p_R)&=\sqrt{2}u(p_R)-w(p_R)\\
\Psi_4(p_R)&=-\sqrt{3}v_t(p_R)
\end{align}
Here $u(p)$ is the s-wave radial wave function, $w(p)$ is the d-wave radial wave function and $v_s(p)$ and $v_t(p)$ are singlet and triplet $p$-wave radial wave functions.

It is convenient to define a half-offshell wave function as
\begin{equation}
\psi_{\lambda_d,s_2}(p_2,P)=G_0(P-p_2)
\Gamma^T_{\lambda_d}(p_2,P)\bar{u}^T(\bm{p}_2,s_2)\,.
\end{equation}
We choose to normalize this wave function such that in the deuteron in any frame
\begin{equation}
\sum_{s_2}\int\frac{d^3p_2}{(2\pi)^3}\frac{m}{E_{p_2}}\bar{\psi}_{\lambda_d,s_2}(p_2,P)
\gamma^\mu \psi_{\lambda_d,s_2}(p_2,P)=\frac{P^\mu}{M_d}\,,
\end{equation}
which is correct only in the absence of energy-dependent kernels.
This results in the normalization of the radial wave functions in the deuteron rest frame being
\begin{align}
1=&\int_0^\infty
\frac{dpp^2}{(2\pi)^3}\left[u^2(p)+w^2(p)+v_t^2(p)+v_s^2(p)\right]\nonumber\\
=&\frac{1}{3}\int_0^\infty
\frac{dpp^2}{(2\pi)^3}\left[\Psi_1^2(p)+\Psi_2^2(p)+\Psi_3^2(p)+\Psi_4^2(p)\right]
\end{align}

In the deuteron rest frame we choose the four-momenta such that
\begin{equation}
p_1=p_p=(E_{p_1},\bm{p}_1)
\end{equation}
\begin{equation}
p_2=(E_p,-\bm{p})
\end{equation}
\begin{equation}
P=(M_d,\bm{0})
\end{equation}
\begin{equation}
q=(\nu,\bm{q})
\end{equation}
The four-momentum of the struck nucleon is given by
\begin{equation}
l=P-p_2=(M_d-E_p,\bm{p})=(E_p,\bm{p})+(M_d-2E_p,\bm{0})=p+\Delta,
\end{equation}
where
\begin{equation}
p=(E_p,\bm{p})
\end{equation}
is onshell and
\begin{equation}
p_R=|\bm{p}|\,.
\end{equation}
The offshell contribution to the momentum of struck nucleon is
\begin{equation}
\Delta=(M_d-2E_p,\bm{0})=(\delta,\bm{0})\,.
\end{equation}

The PWIA response tensor is then
\begin{align}
W^{\mu\nu}_{aa}=&\sum_{s_1,s_2,\lambda_d,\lambda_{d'}}\overline{\psi}_{\lambda_{d'},s_2}(p_2,P)\Gamma^\mu(-q)u(\bm{p}_1,s_1)\bar{u}(\bm{p}_1,s_1)\Gamma^\nu(q)\psi_{\lambda_d,s_2}(p_2,P)\rho_{\lambda_d,\lambda_{d'}}\nonumber\\
=&{\rm Tr}[\Gamma^\mu(-q)\Lambda_+(\bm{p}_1)\Gamma^\nu(q)N(p_2,P)]\,,
\end{align}
where the momentum distribution operator is given by
\begin{equation}
N(p_2,P)=\sum_{s_2,\lambda_d,\lambda_{d'}}\psi_{\lambda_d,s_2}(p_2,P)\rho_{\lambda_d,\lambda_{d'}}\overline{\psi}_{\lambda_{d'},s_2}(p_2,P)\label{eq:momentum_op}
\end{equation}

The deuteron density matrix can be written as
\begin{equation}
\bm{\rho}^D=\frac{1}{3}\left[\sum_{J=0}^2T_{J0}\bm{\tau}_{J0}
+\sum_{J=1}^2\sum_{M=1}^J\left(\Re(T_{JM})\bm{\tau}^\Re_{JM}+\Im(T_{JM})\bm{\tau}^\Im_{JM}\right)\right]\,.\label{eq:density_matrix}
\end{equation}
where $T_{00}=1$ and the matrices $\bm{\tau}_{JM}$ are defined as
\begin{eqnarray}
\bm{\tau}_{00}=\left(
\begin{array}{ccc}
1 & 0 & 0\\
0 & 1 & 0\\
0 & 0 & 1
\end{array}
\right),&
\bm{\tau}_{10}=\sqrt{\frac{3}{2}}\left(
\begin{array}{ccc}
1 & 0 & 0\\
0 & 0 & 0\\
0 & 0 & -1
\end{array}
\right),&
\bm{\tau}_{20}=\frac{1}{\sqrt{2}}\left(
\begin{array}{ccc}
1 & 0 & 0\\
0 & -2 & 0\\
0 & 0 & 1
\end{array}
\right),\nonumber\\
\bm{\tau}_{11}^{\Re}=\sqrt{\frac{3}{2}}\left(
\begin{array}{ccc}
0 & -1 & 0\\
-1 & 0 & -1\\
0 & -1 & 0
\end{array}
\right),&
\bm{\tau}_{11}^{\Im}=\sqrt{\frac{3}{2}}\left(
\begin{array}{ccc}
0 & i & 0\\
-i & 0 & i\\
0 & -i & 0
\end{array}
\right),&\nonumber\\
\bm{\tau}_{21}^{\Re}=\sqrt{\frac{3}{2}}\left(
\begin{array}{ccc}
0 & -1 & 0\\
-1 & 0 & 1\\
0 & 1 & 0
\end{array}
\right),&
\bm{\tau}_{21}^{\Im}=\sqrt{\frac{3}{2}}\left(
\begin{array}{ccc}
0 & i & 0\\
-i & 0 & -i\\
0 & i & 0
\end{array}
\right),&\nonumber\\
\bm{\tau}_{22}^{\Re}=\sqrt{3}\left(
\begin{array}{ccc}
0 & 0 & 1\\
0 & 0 & 0\\
1 & 0 & 0
\end{array}
\right),&
\bm{\tau}_{22}^{\Re}=\sqrt{3}\left(
\begin{array}{ccc}
0 & 0 & -i\\
0 & 0 & 0\\
i & 0 & 0
\end{array}
\right)&\label{taumatrices}
\end{eqnarray}
Note that these matrices are all hermitian. The polarization coefficients can be extracted from the density matrix using
\begin{align}
T_{J0}=&\mathrm{Tr}[\bm{\tau}^\Re_{JM}\bm{\rho}^D]\\
\Re(T_{JM})=&\frac{1}{2} \mathrm{Tr}[\bm{\tau}^\Re_{JM}\bm{\rho}^D]\\
\Im(T_{JM})=&\frac{1}{2} \mathrm{Tr}[\bm{\tau}^\Im_{JM}\bm{\rho}^D]\label{eq:projections}
\end{align}

Using (\ref{eq:density_matrix}), (\ref{taumatrices}) and (\ref{eq:projections}),the momentum distribution operator can be written as
\begin{equation}
N(p_2,P)=\sum_{J=0}^2 N_{J0}(p_2,P)T_{J0}
+\sum_{J=1}^2\sum_{M=1}^J\left[\Re(N_{JM}(p_2,P))\Re(T_{JM})+\Im(N_{JM}(p_2,P))\Im(T_{JM})\right]
\end{equation}
This is an operator in the four-dimensional Dirac spinor space and can be expanded in terms of gamma matrices such that for $J=0$ or $J=2$,
\begin{equation}
N_{JM}(p_2,P)=\frac{1}{8\pi}\left[{\cal N}_{tv}(p_2,P)n_{tv}^{JM}(\bm{p})+{\cal N}_{sv}(P,p2)n_{sv}^{JM}(\bm{p})+{\cal N}_{s}(p_2,P)n_s^{JM}(\bm{p})\right]
\end{equation}
where
\begin{align}
{\cal N}_{tv}(p_2,P)=&\frac{1}{2}\frac{P\cdot p_2}{M_d^2m_N}\gamma\cdot P\\
{\cal N}_{sv}(p_2,P)=&\frac{1}{2}\left(\frac{P\cdot p_2}{M_d^2m_N}\gamma\cdot P-\frac{\gamma\cdot p_2}{m_n}\right)\\
{\cal N}_{s}(p_2,P)=&\frac{1}{2}
\end{align}
and for $J=1$
\begin{equation}
N_{1M}(p_2,P,s)=\frac{1}{8\pi}\left[{\cal N}_{tav}(p_2,P,s)n_{tav}^{1M}(\bm{p})+{\cal N}_{sav}(p_2,P,s)n_{sav}^{1M}(\bm{p})+{\cal N}_{at}(P,p_2,s)n_{at}^{1M}(\bm{p})\right]\,.
\end{equation}
where
\begin{align}
{\cal N}_{tav}(p_2,P,s)=&\frac{1}{2}\frac{P\cdot s}{M_d^2}\gamma\cdot P\gamma_5\\
{\cal N}_{sav}(p_2,P,s)=&\frac{1}{2}\left(\gamma\cdot s-\frac{P\cdot s}{M_d^2}\gamma\cdot P\right)\gamma_5\\
{\cal N}_{at}(p_2,P,s)=&-\frac{i}{2}\frac{m_N}{P\cdot p_2}\sigma^{\alpha\beta}P_\alpha s_\beta\gamma_5
\end{align}
and
\begin{equation}
s=\left(\frac{|\bm{p}|}{m_N},\frac{E_p}{m_N} \hat{p}\right)
\end{equation} is the spin-four vector for rest-frame spin aligned along $\hat{p}$.

The nine momentum distributions are given by
\begin{align}
n_{tv}^{00}(p)=&\frac{1}{3}
   \left(\Psi_1^2(p)+\Psi_2^2(p)+\Psi_3^2(p)+\Psi_4^2(p)\right)\\
n_{sv}^{00}(p)=&\frac{1}{3}\Bigl(\left(\Psi_1^2(p)-\Psi_2^2(p)+\Psi_3^2(p)-\Psi_4^2(p)\right)\nonumber\\
   &+2 \frac{m_N}{p} (\Psi_1(p)
         \Psi_2(p)-\Psi_3(p)
         \Psi_4(p))\Bigr)\\
n_s^{00}(p)=&\frac{1}{3}\Bigl(
   \left(\Psi_1^2(p)-\Psi_2^2(p)+\Psi_3^2(p)-\Psi_4^2(p)\right)\nonumber\\
   &+\frac{p}{m_N}
   (2 \Psi_3(p) \Psi_4(p)-2 \Psi_1(p)
   \Psi_2(p))\Bigr)\nonumber\\
n_{tav}^{1M}(\bm{p})=&-\frac{\eta_M}{3}\sqrt{2 \pi }  \left(\left(\Psi_3^2(p)-\Psi_4^2(p)\right)-2
   \frac{m_N}{p} \Psi_3(p)
   \Psi_4(p)
   \right)Y_{1M}(\Omega_p)\\
n_{sav}^{1M}(\bm{p})=&-\frac{\eta_M}{3} \sqrt{2 \pi }
   \left(\Psi_3^2(p)+\Psi_4^2(p)\right)Y_{1M}(\Omega_p)\\
n_{at}^{1M}(\bm{p})=&-\frac{\eta_M}{3}\sqrt{2 \pi }
   \left(
   \left(\Psi_3^2(p)-\Psi_4^2(p)\right)
   +2 \frac{p}{m_N} \Psi_3(p)
   \Psi_4(p)\right)Y_{1M}(\Omega_p)\nonumber\\
n_{tv}^{2M}(\bm{p})=&-\frac{\eta_M}{3} \sqrt{\frac{2 \pi }{5}}
    \left(2 \Psi_1^2(p)+2
   \Psi_2^2(p)-\Psi_3^2(p)-\Psi_4^2(p)\right)Y_{2M}(\Omega_p)\\
n_{sv}^{2M}(\bm{p})=&-\frac{\eta_M}{3}\sqrt{\frac{2 \pi }{5}}
   \Bigl(\left(2
      \Psi_1^2(p)-2
      \Psi_2^2(p)-\Psi_3^2(p)+\Psi_4^2(p)
      \right)\nonumber\\
         &+2 \frac{m_N}{p} (2 \Psi_1(p)
            \Psi_2(p)+\Psi_3(p)
            \Psi_4(p)) \Bigr)Y_{2M}(\Omega_p)\\
n_{s}^{2M}(\bm{p})=&-\frac{\eta_M}{3}\sqrt{\frac{2 \pi }{5}}\Bigl( \left(2 \Psi_1^2(p)-2\Psi_2^2(p)-\Psi_3^2(p)+\Psi_4^2(p)\right)\nonumber\\
&-2 \frac{p}{m_N} (2 \Psi_1(p)\Psi_2(p)+\Psi_3(p)\Psi_4(p))\Bigr) Y_{2M}(\Omega_p)
\end{align}
where
\begin{equation}
\eta_M=\left\{
\begin{array}{lr}
1 &{\rm for}\ M=0\\
2 &{\rm for}\ M>0
\end{array}\right.\,.
\end{equation}

If the $p$-waves are neglected then $\Psi_2\rightarrow0$ and $\Psi_4\rightarrow0$. The momentum distributions then simplify to
\begin{equation}
n_+^{00}=n_{tv}^{00}(p)=n_{sv}^{00}(p)=n_s^{00}(p)=\frac{1}{3}
   \left(\Psi_1^2(p)+\Psi_3^2(p)\right)
\end{equation}
\begin{equation}
n_+^{1M}(\bm{p})=n_{tav}^{1M}(\bm{p})=n_{sav}^{1M}(\bm{p})=n_{at}^{1M}(\bm{p})=-\frac{\eta_M}{3} \sqrt{2 \pi } \Psi_3^2(p)
   Y_{1M}(\Omega_p)
\end{equation}
\begin{equation}
n_+^{2M}(\bm{p})=n_{tv}^{2M}(\bm{p})=n_{sv}^{2M}(\bm{p})=n_{s}^{2M}(\bm{p})=-\frac{\eta_M}{3} \sqrt{\frac{2 \pi }{5}}
    \left(2
      \Psi_1^2(p)-\Psi_3^2(p)\right)Y_{2M}(\Omega_p)\,.
\end{equation}
These can be rewritten in terms of $u$ and $w$ using
\begin{align}
&\frac{1}{3}
   \left(\Psi_1^2(p)+\Psi_3^2(p)\right)=u^2(p)+w^2(p)\\
&\frac{1}{3}\Psi_3^2(p)=\frac{1}{3} \left(2 u(p)^2-2 \sqrt{2} u(p)
   w(p)+w(p)^2\right)\\
&\frac{1}{3}\left(2\Psi_1^2(p)-\Psi_3^2(p)\right)=w(p) \left(2 \sqrt{2} u(p)+w(p)\right)\,.
\end{align}
These are in agreement with the usual nonrelativistic polarized momentum distributions up to a factor determined by our choice for the normalization of the wave functions \cite{Frankfurt:1983qs,Bianconi}.

Since all of the momentum distributions are now the same for each $J$ and $M$, these can now be factored out and leave the combinations of operators
\begin{equation}
{\cal N}_{tv}(p_2,P)+{\cal N}_{sv}(p_2,P)+{\cal N}_{s}(p_2,P)=\Lambda_+(\bm{p})
\end{equation}
for $J=0,2$ and
\begin{equation}
{\cal N}_{tav}(p_2,P)+{\cal N}_{sav}(p_2,P)+{\cal N}_{at}(p_2,P)=\frac{1}{2}\left[\gamma\cdot s-\frac{i}{2}\frac{m_N}{P\cdot p_2}\sigma^{\alpha\beta}P_\alpha s_\beta\right]\gamma_5
\end{equation}
for $J=1$.

\begin{align}
N(p_2,P)=&\frac{1}{8\pi}\Biggl\{\Lambda_+(\bm{p})\left[n_+^{00}(p)+n_+^{20}(\bm{p})T_{20}+\sum_{M=1}^2\left[\Re(n_+^{2M}(\bm{p}))\Re(T_{2M})+\Im(n_+^{2M}(\bm{p})))\Im(T_{2M})\right]\right]\nonumber\\
&+\frac{1}{2}\left[\gamma\cdot s-\frac{i}{2}\frac{m_N}{P\cdot p_2}\sigma^{\alpha\beta}P_\alpha s_\beta\right]\gamma_5\nonumber\\
&\times\left[n_+^{10}(\bm{p})T_{20}+\Re(n_+^{11}(\bm{p}))\Re(T_{11})+\Im(n_+^{11}(\bm{p}))\Im(T_{11}))\right]\Biggr\}
\end{align}

The factored cross section can then be written as
\begin{align}
\frac{ d \sigma^5}{d \epsilon' d \Omega_e d \Omega_p} =&  \frac{m_p \, m_n \, p_1}{8 \pi^3 \, M_d} \,
\sigma_{Mott} \,
f_{rec}^{-1} \,
\Biggl\{ [  v_L r_L^{(I)} +   v_T r_T^{(I)}
 + v_{TT} r_{TT}^{(I)}\cos 2\phi+ v_{LT}r_{LT}^{(I)}\cos\phi
 ]\nonumber\\
 & \left[n_+^{00}(p)+n_+^{20}(\bm{p})T_{20}+\sum_{M=1}^2\left[\Re(n_+^{2M}(\bm{p}))\Re(T_{2M})+\Im(n_+^{2M}(\bm{p}))\Im(T_{2M})\right]\right]\nonumber\\
 &+h[ v_{LT'}r_{LT'}^{(II)}\cos\phi+v_{T'}r_{T'}^{(II)}]\nonumber\\
 &\times\left[n_+^{10}(\bm{p})T_{10}+\Re(n_+^{11}(\bm{p}))\Re(T_{11})+\Im(n_+^{11}(\bm{p}))\Im(T_{11})\right]\Biggr\}
\label{def5sigma}\,,
\end{align}
where the effective reduced single-nucleon response functions are listed in Appendix \ref{app:single_nucleon} and are related to the conventional deForest CC1 prescription \cite{DeForest:1983vc} up to normalization factors. Note that contributions from vector polarization of the deuteron contribute to the factored cross section only when the electron beam is also polarized. The single-nucleon contributions for unpolarized electrons are the same for both unpolarized and tensor polarized deuterons.

Assuming that $\widetilde{T}_{JM}\neq 0$ only for $J=2$ and $M=0$ and using (\ref{eq:T_Ttilde}) the tensor asymmetry for the factored cross section can be written as
\begin{align}
(A^T_d)_{factored}=&\frac{n_+^{20}(\bm{p})T_{20}+\Re(n_+^{21}(\bm{p}))\Re(T_{21})+\Re(n_+^{22}(\bm{p}))\Re(T_{22})}{n_+^{00}(p)\widetilde{T}_{20}}\nonumber\\
=&\frac{n_+^{20}(\bm{p})\frac{1}{4}(1+3\cos 2\theta_{kq})
	+\Re(n_+^{21}(\bm{p}))\sqrt{\frac{3}{8}}\sin 2
	\theta_{kq}+\Re(n_+^{22}(\bm{p}))\sqrt{\frac{3}{32}}
	(1-\cos 2\theta_{kq})}{n_+^{00}(p)}\nonumber\\
=&- \sqrt{\frac{2 \pi }{5}}
\frac{2
	\Psi_1^2(p)-\Psi_3^2(p)}{
	\Psi_1^2(p)+\Psi_3^2(p)}\Xi(\theta_m,\phi,\theta_{kq})
\end{align}
where the factored effective single-proton cross section is canceled since it is the same for both the numerator and denominator. The angular factor is defined as
\begin{align}
\Xi(\theta_m,\phi,\theta_{kq})=&\left[\frac{1}{4}(1+3\cos 2\theta_{kq})Y_{20}(\Omega_p)
+\sqrt{\frac{3}{2}}\sin 2
\theta_{kq}\Re(Y_{21}(\Omega_p))\right.\nonumber\\
&\left.+\sqrt{\frac{3}{8}}
(1-\cos 2\theta_{kq})\Re(Y_{22}(\Omega_p)) \right]\nonumber\\
=&\sqrt{\frac{5}{64\pi}}\left[\frac{1}{4}(1+3\cos 2\theta_{kq})(1+3\cos 2\theta_m)-3\sin 2\theta_{kq}\sin 2\theta_m\cos\phi\right.\nonumber\\
&\left.+\frac{3}{4}(1-\cos\theta_{kq})(1-\cos\theta_m)\cos 2\phi\right]\,.
\end{align}
If we define a reduced tensor asymmetry as
\begin{equation}
a^T_d=\frac{A^T_d}{\Xi(\theta_m,\phi,\theta_{kq})}
\end{equation}
Then the factored reduced tensor asymmetry is
\begin{equation}
{(a^T_d)}_{factored}=- \sqrt{\frac{2 \pi }{5}}
\frac{w(p)(2\sqrt{2}u(p)+w(p))}{u^2(p)+w^2(p)}
\label{atd_factored}
\end{equation} 
independent of all kinematical variables except the missing momentum $p$.

 In Section \ref{secresults}, we 
investigate numerically the behavior of the reduced asymmetry when it is calculated using final state interactions,
and various versions of commonly used wave functions and form factor parameterizations. We will observe there that
factorization breaks down, and that the use of (\ref{atd_factored}) is unrealistic.

\section{Results}
\label{secresults}

Independent of the dynamical model for the description of the $\vec H^2 (e,e'p)$ reaction,
every calculation needs a wave function, nucleon form factor parametrizations, and 
nucleon-nucleon amplitudes as inputs. We list the model inputs used in our calculations
in Table \ref{tab:model_inputs}. The reasons for these choices are discussed in more detail in \cite{Ford:2014yua}.

\begin{table}\caption{Model inputs to the calculation.}
 \begin{center}
\begin{tabular}{ c | c | c  } 
\hline
Final State Interactions  & Form Factors & Deuteron Wave Function \\
\hline \hline
\multirow{8}{4em}{Regge \cite{FVO_Reggemodel,FJVO,Ford:2013wxa} \\SAID \cite{SAIDdata,Arndt07,Arndt00} } & \multirow{8}{4em}{GKex05\cite{FF_GK_05_2_Lomon_02,FF_GK05_1_Lomon_06} \\ AMT\cite{FF_AMT_07} \\ MMD\cite{FF_MMD1_Mergell_96}} & IIB \cite{Gross:1991pm}\\ 
			 &  & WJC 1\cite{Gross:2007jj} \\ 
                         &  & WJC 2\cite{Gross:2007jj}  \\ 
                         &  & AV18 \cite{Wiringa95} \\ 
			&  & CD Bonn \cite{Machleidt:2000ge}\\ 
                               &  & NIMJ 1 \cite{Stoks:1994wp}\\ 
                                &  & NIMJ 2\cite{Stoks:1994wp} \\ 
                                &  & NIMJ 3\cite{Stoks:1994wp} \\ 
\hline
\end{tabular}
\end{center} \label{tab:model_inputs}
\end{table}

\begin{figure}
\centerline{\includegraphics[height=3in]{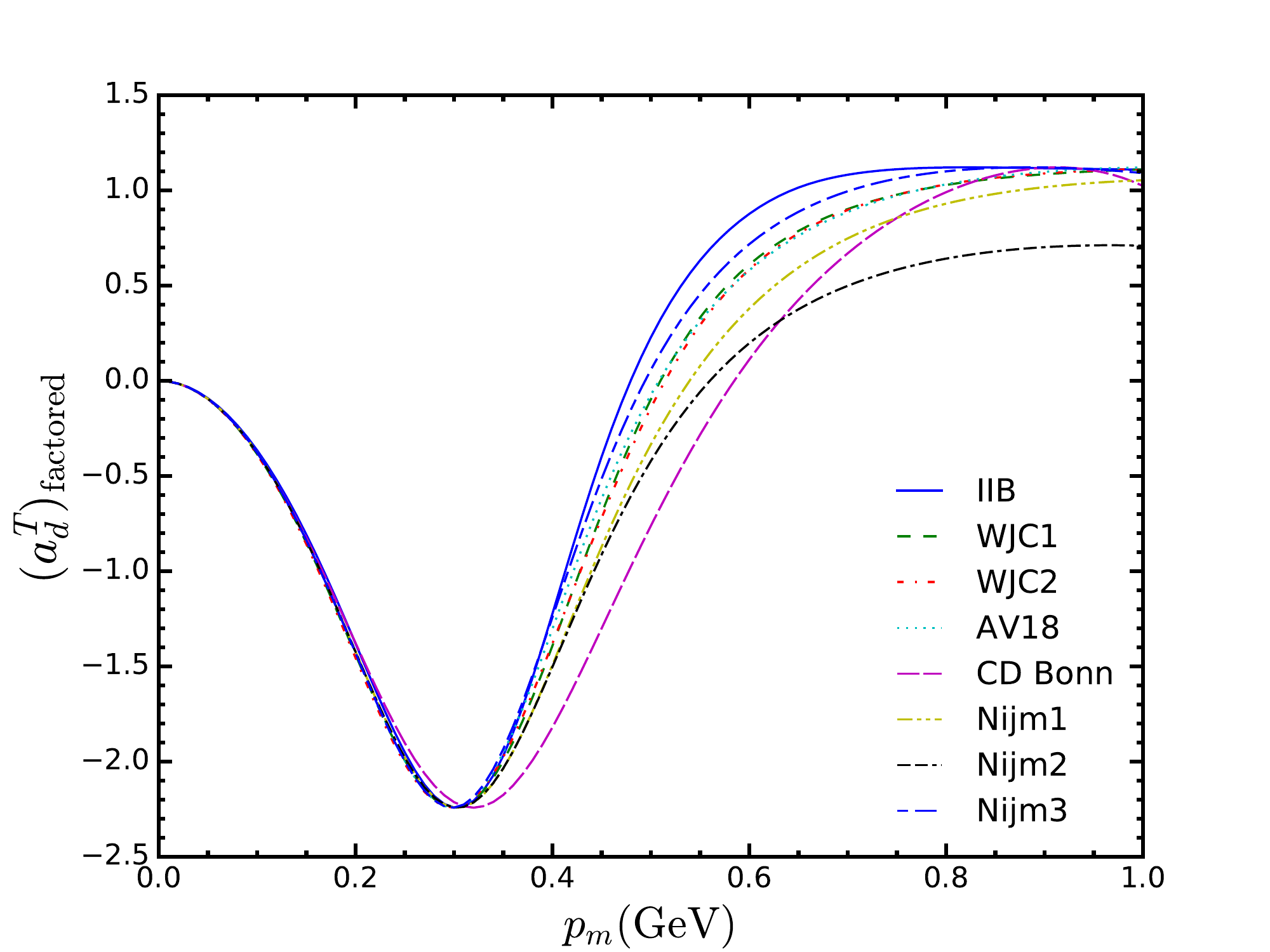}}
\caption{(color online) The factored, reduced asymmetry $a^T_d$ calculated for the eight different wave functions
used in our calculations.  }\label{fig_atd_pwia}
\end{figure}

We start out by showing the influence of the different wave functions on the factored, reduced asymmetry $a^T_d$
as defined in eqn. \ref{atd_factored} in 
Fig. \ref{fig_atd_pwia}. 
Note that the factored, reduced asymmetry does not depend on the nucleon form factors, so the only model input necessary
is the wave function.
For missing momenta larger than $0.3$ GeV, the curves start to deviate from each other, and they fan out
considerably for $p_m \approx 0.6$ GeV and larger. Most wave functions lead to similar asymmetries, with the 
exception of the CD Bonn that has a slightly different shape, peaking at high $p_m$, and the Nijmegen 3 wave function
that leads to the lowest values for the factored, reduced $a^T_d$ at the high $p_m$.

\begin{table}
	\caption{Wave function probabilities.}
	\begin{tabular}{c|c|c|c|c}
		\hline\hline
		&s wave	&d wave	&triplet p wave &singlet p wave \\\hline
		IIB &	94.82\% &	5.11\% &	0.06\% &	0.01\%\\
		WJC 1 &	92.33\% &	7.34\% &	0.11\% &	0.21\%\\
		WJC 2 &	93.60\% &	6.38\% &	0.01\% &	0.01\%\\
		AV18 &	94.24\% &	5.76\% &	0.00\%  &	0.00\%\\
		CD Bonn	&	95.15\% &	4.85\% &	0.00\% &	0.00\%\\
		NIJM 1	&	94.25\% &	5.75\% &	0.00\% &	0.00\%\\
		NIJM 2	&	94.32\% &	5.68\% &	0.00\% &	0.00\%\\
		NIJM 3	&	94.35\% &	5.65\% &	0.00\% &	0.00\%\\
		\hline\hline
	\end{tabular}
	\label{tab_wf_contents}
\end{table}

When comparing the results shown in Fig. \ref{fig_atd_pwia} to the numbers for the $d$ wave content of the various
wave functions in Table \ref{tab_wf_contents}, it is obvious that there is no direct relationship between the $d$-wave
content and the size of the reduced asymmetry calculated with a particular wave function. Depending on the missing momentum, e.g. the CD Bonn wave function result is either below the others (for $p_m < 0.6$ GeV), or above the
others around $p_m \approx 0.9$ GeV. CD Bonn has the lowest $d$ state probability of all considered wave functions. Nijmegen 2 and Nijmegen 3 have almost identical $d$ wave contents - $5.68 \%$ versus $5.65 \%$ - but are rather different, with Nijmegen 2 leading to a much smaller asymmetry than Nijmegen 3 for $p_m > 0.5$ GeV.

A more realistic calculation includes the Born approximation graph, where the photon couples to the neutron,
and final state interactions. At this point, the parametrizations of the nucleon form factors and of the
nucleon-nucleon amplitudes enter. As eight wave functions, three form factor parametrizations,
and two nucleon-nucleon amplitude parametrizations lead to $8 \times 3 = 24$ possible combinations for PWBA
and to $8 \times 3 \times 2 = 48$ possible combinations for the DWBA (henceforth referred to as FSI) and therefore
lead to very busy
plots, we only show the envelope of the PWBA and FSI calculations in the figures.

We remark in passing that the differences between PWIA and PWBA calculations for the same choice of model inputs
is tiny. The difference is apparent in the numbers, but does not show up on a plot of the scale we use for the figures in this paper. The use of different form factor parametrizations in PWBA leads to a relatively larger, but absolutely still very small difference that is not visible at the scale used. 

The difference between the PWBA and the factored version of the PWBA, which excludes $p$-waves, is small but visible
at medium and high missing momentum when plotted for relativistic wave functions. The non-relativistic wave functions
still show a difference between factored and unfactored PWBA, but this is tiny as it is practically the difference
between PWIA and PWBA.

\begin{figure}
 \centerline{\includegraphics[height=3in]{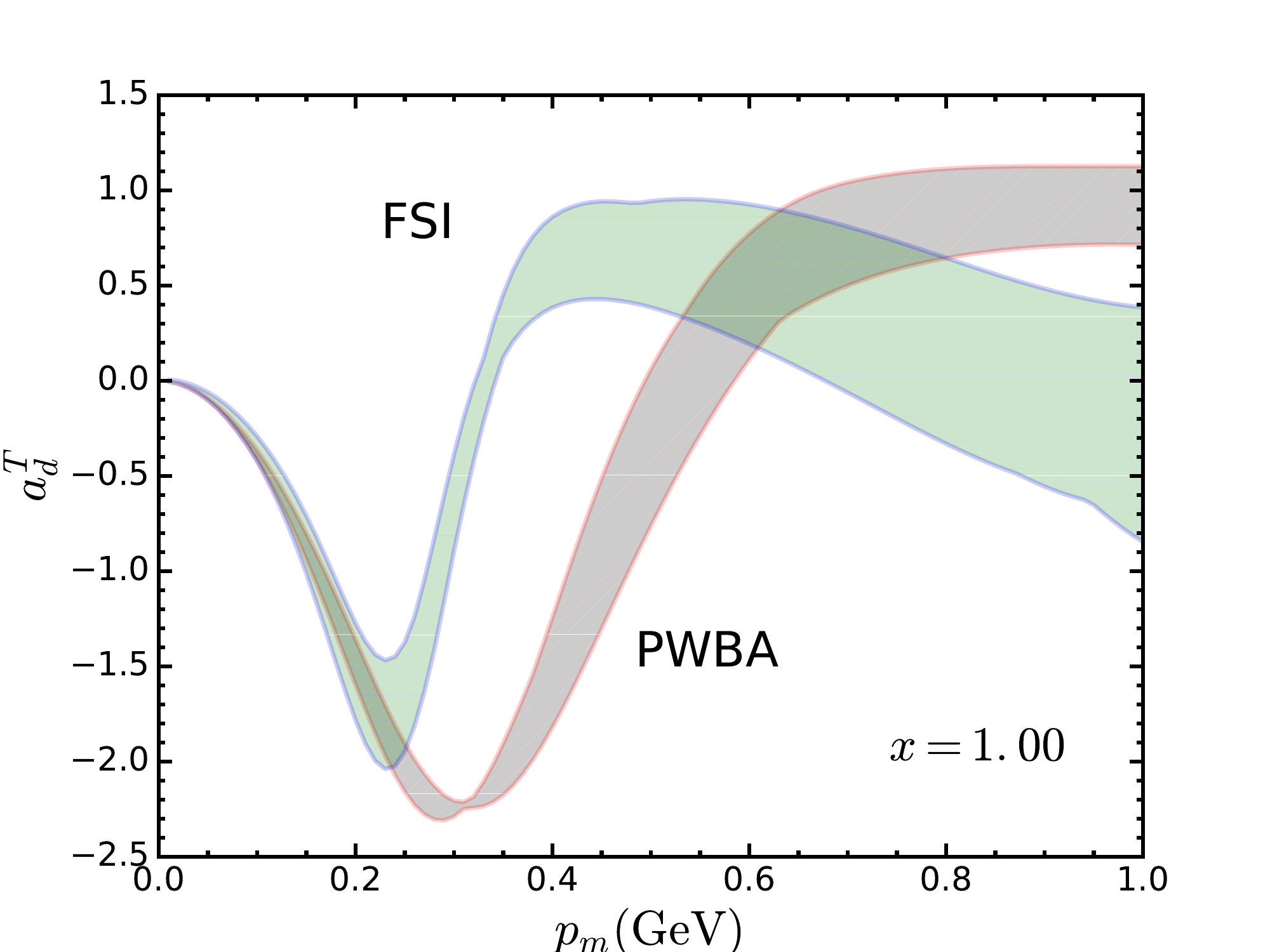}}
 \caption{(color online) The envelopes of the reduced asymmetry $a^T_d$ calculated in PWBA and with FSI, 
 for $x = 1.00$, $Q^2=2.4\ \mathrm{GeV^2}$, $\varepsilon=8.3\ \mathrm{GeV}$ and $\phi=165^\circ$.   }\label{fig_atd_both_x1}
\end{figure}

In Fig. \ref{fig_atd_both_x1} we show the envelopes for the PWBA and FSI calculations for $x = 1.00$. The differences
in the PWBA calculations mainly stem from the different wave functions used, and the PWBA envelope shown
mainly corresponds to the PWIA curves of Fig. \ref{fig_atd_pwia}. Once FSIs are included, the asymmetry changes
shape, and the dip moves to lower values of the missing momentum, as observed already in \cite{targetpol}.
For FSI, the different model inputs now lead to a significant spread for missing momenta above $0.35$ GeV, as well
as in the dip of the asymmetry at lower $p_m$. For high missing momentum, the uncertainties introduced by the 
model inputs in FSI are more than twice as large as for PWBA.

\begin{figure}
 \centerline{\includegraphics[height=3in]{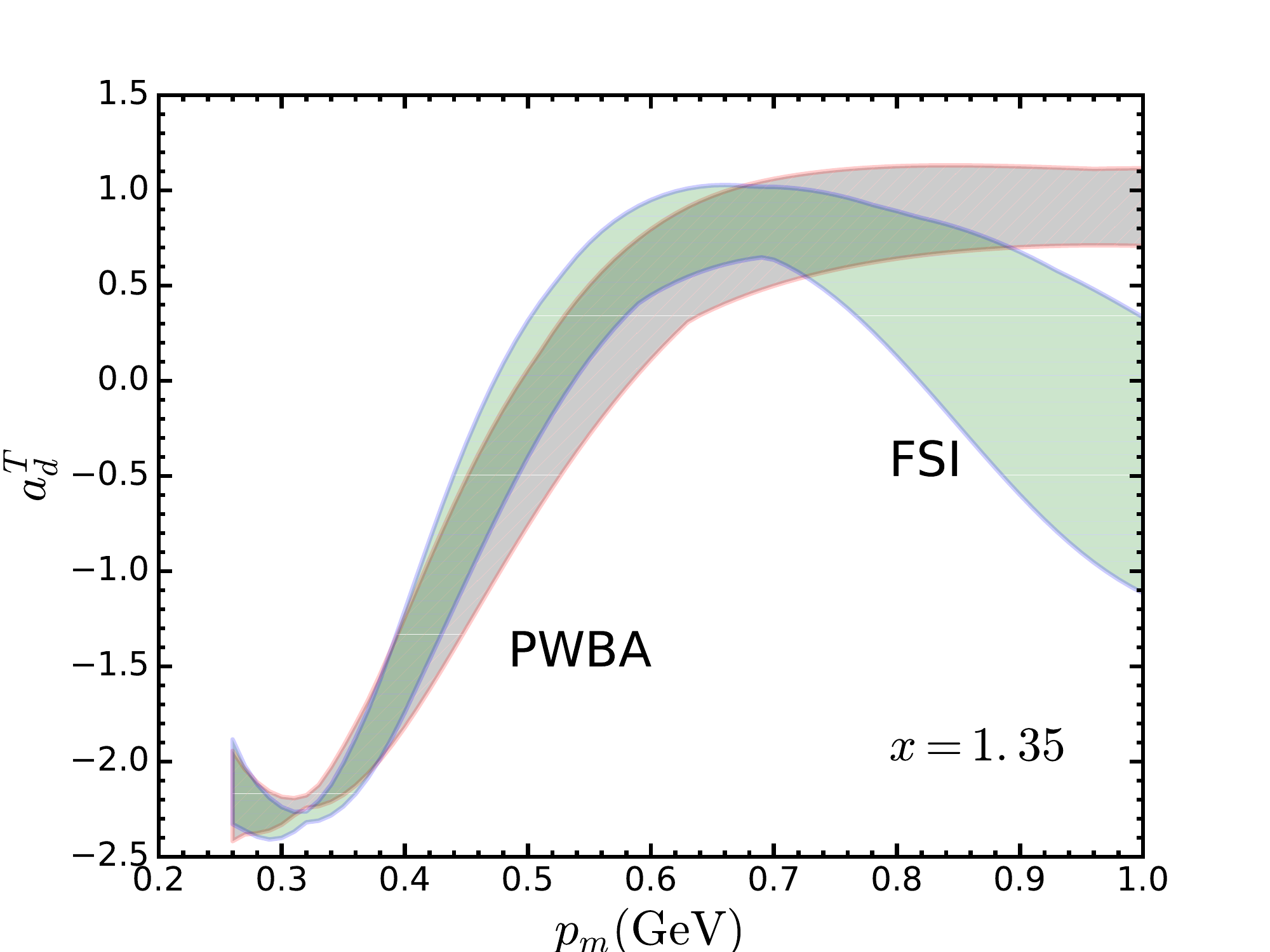}}
 \caption{(color online) The envelopes of the reduced asymmetry $a^T_d$ calculated in PWBA and with FSI, 
 for $x = 1.35$, $Q^2=4.25\ \mathrm{GeV^2}$, $\varepsilon=11\ \mathrm{GeV}$ and $\phi=165^\circ$.   }\label{fig_atd_both_x135}
\end{figure}

We now consider kinematics at $x = 1.35$, away from the quasi-elastic peak. Our results are shown in Fig. \ref{fig_atd_both_x135}. As for $x = 1.00$, the FSI calculation envelope shows a much larger spread due to the
model inputs than the PWBA envelope, in particular for missing momenta larger than $0.7$ GeV.
While in the factored version of the calculation, i.e. in PWIA without $p$-waves, the results are completely 
independent of $x$, it is obvious from comparing Fig. \ref{fig_atd_both_x1} and Fig. \ref{fig_atd_both_x135} that
FSIs introduce a quite drastic dependence on kinematic variables beyond the missing momentum. 

\section{Summary and Outlook}
 
In this paper, first we considered the tensor asymmetry $A^T_d$ within a relativistic framework. We investigated
the conditions under which this asymmetry can be factored. We defined a reduced asymmetry $a^T_d$
that factors in PWIA when the $p$-waves are neglected. The factored version of the reduced asymmetry
depends on the missing momentum only. It agrees with  the 
well-known non-relativistic version.

Then, we presented numerical results for the reduced tensor asymmetry $a^T_d$ and compared
these results to the factored version  $(a_T^d)_{factored}$.  
While we have shown analytically and numerically that factorization holds
in PWIA in the absence of $p$-waves, the numerical results imply that factorization is broken thoroughly 
once FSIs are included. The inclusion of FSIs leads to changes in shape of the reduced asymmetry, in
particular at high missing momentum. The FSIs also introduce a significant dependence on $x$ (and other kinematics
variables)
besides the missing momentum, thus making the breaking of factorization obvious. This is consistent with
\cite{sabine_factor,targetpol}.

We also have demonstrated numerically that find that 
there is no systematic relationship of form of $(a^T_d)_{factored}$ to the d-wave probability of the  various wave functions.  

Our results imply that extracting any information on the $d$-wave content of the wave 
function - which is, after all, not an observable - from the tensor asymmetry $A^T_d$ will require an extremely careful treatment of the reaction dynamics, and will carry
a theoretical uncertainty due to the many, equally valid model inputs necessary.
The kinematics at large $x$ and medium values of $p_m$ might be best suited to any such attempt.

{\bf Acknowledgments}: This work was
supported in part by funds provided by the U.S. Department of Energy
(DOE) under cooperative research agreement under No.
DE-AC05-84ER40150 and by the National Science Foundation under
grant No. PHY-1306250.

\appendix

\section{kinematic factors and response functions}\label{app:kinematic}

The leptonic coefficients $v_K$ are
\begin{eqnarray}
v_L&=&\frac{Q^4}{q^4}\\
v_T&=&\frac{Q^2}{2q^2}+\tan^2\frac{\theta_e}{2}\\
v_{TT}&=&-\frac{Q^2}{2q^2}\\
v_{LT}&=&-\frac{Q^2}{\sqrt{2}q^2}\sqrt{\frac{Q^2}{q^2}+\tan^2\frac{\theta_e}{2}}\\
v_{LT'}&=&-\frac{Q^2}{\sqrt{2}q^2}\tan\frac{\theta_e}{2}\\
v_{T'}&=&\tan\frac{\theta_e}{2}\sqrt{\frac{Q^2}{q^2}+\tan^2\frac{\theta_e}{2}}\,.
\end{eqnarray}

The response functions in the xyz-frame are given by
\begin{eqnarray}
R_L(D)&=&R_L^{(I)}(D)=W^{00}(D)\nonumber\\
R_T(D)&=&R_T^{(I)}=W^{11}(D)+W^{22}(D)\nonumber\\
R_{TT}(D)&=&R_{TT}^{(I)}(D)\cos 2\phi+R_{TT}^{(II)}(D)\sin 2\phi\nonumber\\
R_{LT}(D)&=&R_{LT}^{(I)}(D)\cos \phi+R_{LT}^{(II)}(D)\sin \phi\nonumber\\
R_{LT'}(D)&=&R_{LT'}^{(I)}(D)\sin \phi+R_{LT'}^{(II)}(D)\cos \phi\nonumber\\
R_{T'}(D)&=&R_{T'}^{(II)}=-2\Re(W^{12}(D))\,,
\label{respdefxyz}
\end{eqnarray}
where
\begin{align}
R_{TT}^{(I)}(D)\cos 2\phi=& W^{22}(D)-W^{11}(D)\nonumber\\
R_{TT}^{(II)}(D)\sin 2\phi=&2\Re(W^{12}(D)) \nonumber\\
R_{LT}^{(I)}(D)\sin\phi=&2\sqrt{2}\Re(W^{01}(D)) \nonumber\\
R_{LT}^{(II)}(D)\cos\phi=&-2\sqrt{2}\Re(W^{02}(D)) \nonumber\\
R_{LT'}^{(I)}(D)\sin\phi=&-2\sqrt{2}\Im(W^{01}(D)) \nonumber\\
R_{LT'}^{(II)}(D)\cos\phi=&-2\sqrt{2}\Im(W^{02}(D))\,. 
\end{align}

\section{rotations}\label{app:rotations}

The form of the differential cross section given by (\ref{def5sigma}) assumes that the deuteron is polarized relative the direction of the momentum transfer $\bf{q}$. As a practical matter, polarized deuteron targets are generally polarized along the direction of the incident beam parallel to $\bf{k}$. Re-expressing the cross section for polarization to lie along the beam simply requires that the density matrix be rotated from $\bm{q}$ to $\bm{k}$. This involves a rotation of the density matrix through an angle $\theta_{kq}$ where

\begin{equation}
\theta_{kq}=\cos^{-1}\frac{|\bm{k}|-|\bm{k}'|\cos\theta_l}{|\bm{q}|}\,.
\end{equation}

The components of the density matrix are given by the matrix elements of the density matrix operator $\hat{\rho}^d$ as
\begin{equation}
\rho^D_{\lambda_d\lambda'_d}=\left<\lambda_d\left|\hat{\rho}^D\right|\lambda'_d\right>
\end{equation}
The density matrix operator $\hat{\tilde{\rho}}$ Aligned along $\bm{k}$ is obtained by a right handed rotation about the y-axis through the angle $\theta_{kq}$ which can be written as
\begin{equation}
\hat{\tilde{\rho}}^D=\hat{R}(\hat{y},\theta_{kq})\rho^D\hat{R}^{-1}(\hat{y},\theta_{kq})\,.
\end{equation}
The inverse of this expression is
\begin{equation}
\hat{\rho}^D=\hat{R}^{-1}(\hat{y},\theta_{kq})\hat{\tilde{\rho}}^D\hat{R}(\hat{y},\theta_{kq})=\hat{R}(\hat{y},-\theta_{kq})\hat{\tilde{\rho}}^D\hat{R}(\hat{y},\theta_{kq})\,.
\end{equation}
So the matrix element of density operator is related to the rotated operator by
\begin{equation}
\rho^D_{\lambda_d\lambda'_d}=\sum_{\lambda''_d\lambda'''_d}d^1_{\lambda_d\lambda''_d}(-\theta_{kq})\tilde{\rho}^D_{\lambda''_d\lambda'''_d}d^1_{\lambda'''_d\lambda'_d}(\theta_{kq})\,.\label{eq:rotation}
\end{equation}

The density matrix can be related the density matrix polarized relative to $\bm{k}$ can be obtained using (\ref{eq:rotation}) where
\begin{equation}
\bm{d}^1(\theta)=\left(
\begin{array}{ccc}
\frac{1}{2} (1+\cos \theta) &
-\frac{1}{\sqrt{2}}\sin \theta &
\frac{1}{2} (1-\cos \theta) \\
\frac{1}{\sqrt{2}}\sin \theta & \cos
\theta & -\frac{1}{\sqrt{2}}\sin \theta \\
\frac{1}{2} (1-\cos \theta) &
\frac{1}{\sqrt{2}}\sin \theta &
\frac{1}{2} (1+\cos \theta) \\
\end{array}
\right)
\end{equation}
and representing the rotated density matrix by
\begin{equation}
\tilde{\bm{\rho}}^D=\frac{1}{3}\left[\sum_{J=0}^2\tilde{T}_{J0}\bm{\tau}_{J0}
+\sum_{J=1}^2\sum_{M=1}^J\left(\Re(\tilde{T}_{JM})\bm{\tau}^\Re_{JM}+\Im(\tilde{T}_{JM})\bm{\tau}^\Im_{JM}\right)\right]\,.
\end{equation}
The polarization coefficients can then be extracted using the properties of the matrices (\ref{taumatrices}) to give the polarization coefficients in terms of the rotated polarization coefficients yielding
\begin{eqnarray}
T_{10}&=&\cos\theta_{kq}\widetilde{T}_{10}
-\sqrt{2}\sin\theta_{kq}\Re{\widetilde{T}_{11}}\nonumber\\\
\Re(T_{11})&=&\frac{1}{\sqrt{2}}
\sin\theta_{kq}\widetilde{T}_{10}
+\cos\theta_{kq}\Re{\widetilde{T}_{11}}
\nonumber\\
\Im(T_{11})&=&\Im{\widetilde{T}_{11}}\nonumber\\
T_{20}&=&\frac{1}{4}(1+3\cos 2\theta_{kq})
\widetilde{T}_{20}-\sqrt{\frac{3}{2}}\sin 2\theta_{kq}\Re{\widetilde{T}_{21}}+\sqrt{\frac{3}{8}}\left(1-\cos 2\theta_{kq}\right)\Re{\widetilde{T}_{22}}\nonumber\\
\Re(T_{21})&=&\sqrt{\frac{3}{8}}\sin 2
\theta_{kq}\widetilde{T}_{20}+\cos 2\theta_{kq}\Re{\widetilde{T}_{21}}
-\frac{1}{2}\sin 2\theta_{kq}\Re{\widetilde{T}_{22}}
\nonumber\\
\Im(T_{21})&=&\cos \theta_{kq}
\Im{\widetilde{T}_{21}}-
\sin\theta_{kq}\Im{\widetilde{T}_{22}}\nonumber\\
\Re(T_{22})&=&\sqrt{\frac{3}{32}}
(1-\cos 2\theta_{kq})\widetilde{T}_{20}
+\frac{1}{2}\sin 2\theta_{kq}
\Re{\widetilde{T}_{21}}+\frac{1}{4}\left(3+\cos 2\theta_{kq}\right)\Re{\widetilde{T}_{22}}\nonumber\\
\Im(T_{22})&=&\sin \theta_{kq}\Im{\widetilde{T}_{21}}+\cos\theta_{kq}\Im{\widetilde{T}_{22}}\,.\label{eq:T_Ttilde}
\end{eqnarray}

\section{single nucleon offshell response function}\label{app:single_nucleon}

The effective single-nucleon offshell response functions are given by
\begin{align}
r_L^{(I)}=&\frac{1}{64\pi m_N^4}\{-4 F_1^2(Q^2) m_N^2 Q^2 - 8 F_1(Q^2) F_2(Q^2) m_N^2
(\nu^2 + Q^2) + 4 E_p^2 (4 F_1^2(Q^2) m_N^2 \nonumber\\
& +F_2^2(Q^2) Q^2)+ 4 E_p \nu (4 F_1^2(Q^2) m_N^2 +
F_2^2(Q^2) Q^2) + F_2^2(Q^2) (\nu^2 Q^2 -
4 m_N^2 (\nu^2 + Q^2))\nonumber\\
&-2\delta (2 E_p + \nu) (-4 F_1^2(Q^2) m_N^2 +
F_2^2(Q^2) (2 E_p \nu + \nu^2 - Q^2))  \nonumber\\
& +\delta^2[-4 E_p^2 F_2^2(Q^2) + 4 F_1^2(Q^2) m_N^2 -
12 E_p F_2^2(Q^2) \nu + F_2^2(Q^2) (-5 \nu^2 + Q^2)] \nonumber\\
&-4\delta^3 F_2^2(Q^2) (E_p + \nu)-\delta^4F_2^2(Q^2)\}\,,
\end{align}
\begin{align}
r_T^{(I)}=&\frac{1}{64\pi m_N^4}\{4 [4 F_1(Q^2) F_2(Q^2) m_N^2 Q^2 +
F_2^2(Q^2) (2 m_N^2 + p_\perp^2) Q^2 +
2 F_1^2(Q^2) m_N^2 (2 p_\perp^2 + Q^2)]\nonumber\\
&-16\delta F_1(Q^2) (F_1(Q^2) + F_2(Q^2)) m_N^2 \nu+\delta^2(8 E_p^2 F_2^2(Q^2) - 8 F_1^2(Q^2) m_N^2 +
8 E_p F_2^2(Q^2) \nu \nonumber\\
&- 2 F_2^2(Q^2) Q^2)+4\delta^3 F_2^2(Q^2) (2 E_p + \nu)+2\delta^4 F_2^2(Q^2)\}\,,
\end{align}
\begin{align}
r_{TT}^{(I)}=&\frac{-4 p_\perp^2 (4 F_1^2(Q^2) m_N^2 + F_2^2(Q^2) Q^2)}{64\pi m_N^4}\,,
\end{align}
\begin{align}
r_{LT}^{(I)}=&\frac{1}{64\pi m_N^4}4 \sqrt{2}\{ (2 E_p + \nu) p_\perp
(4 F_1^2(Q^2) m_N^2 + F_2^2(Q^2) Q^2)\nonumber\\
&+\delta p_\perp [4 F_1^2(Q^2) m_N^2 +
F_2^2(Q^2) (-2 E_p \nu - \nu^2 + Q^2)]-\delta^2 F_2^2(Q^2) \nu p_\perp
\}\,,
\end{align}

\begin{align}
r_{LT'}^{(II)}=& \frac{ \sqrt{2}p_\perp q}{16\pi m_N^4p}[
E_p^2 F_2^2(Q^2) \nu -4 E_p F_1(Q^2)
m_N^2 (F_1(Q^2)+F_2(Q^2)\nonumber\\
&+F_2(Q^2) \nu  \left(2
F_1(Q^2) m_N^2+F_2(Q^2)\right)
\left(m_N^2-p^2\right)\nonumber\\
&+\delta E_p F_2^2(Q^2)   (2
E_p+\nu )+\delta^2 E_p F_2^2(Q^2) ]
\end{align}

\begin{align}
r_{T'}^{(II)}=&\frac{1}{16\pi m_N^4p}\Biggl[ \Bigl(E_p^2 p_\parallel \left(4 F_1^2(Q^2)
m_N^2+F_2^2(Q^2) Q^2\right)+4 E_p
F_1(Q^2) m_N^2 \nu  p_\parallel
(F_1(Q^2)+F_2(Q^2))\nonumber\\
&-4 F_1^2(Q^2) m_N^2
\left(m_N^2 p_\parallel+p^2
(p_\parallel+q)\right)+2 F_1(Q^2) F_2(Q^2) m_N^2
\left(p_\parallel Q^2-2 p^2 q\right)\nonumber\\
&+F_2^2(Q^2)
p_\parallel Q^2
\left(m_N^2-p^2\right)\Bigr)+\delta E_p  \left(F_2^2(Q^2) \left(-2 E_p \nu
p_\parallel+2 p^2 q+p_\parallel Q^2\right)\right.\nonumber\\
&+\left.
4F_1^2(Q^2) m_N^2 p_\parallel+ 4F_1(Q^2) F_2(Q^2)
m_N^2 p_\parallel\right)\nonumber\\
&+\delta^2  F_2(Q^2) \left(-E_p F_2(Q^2) \nu
p_\parallel+2 F_1(Q^2) m_N^2 p_\parallel+F_2(Q^2)
p^2 q\right)\Biggr]
\end{align}
where  $p_\perp=p\sin\theta_{m}$ and $p_\parallel=p\cos\theta_{m}$.

\bibliography{pol_mom_dist}

\end{document}